\documentclass{emulateapj}
\shorttitle{Metallicities of UCDs}
\shortauthors{Mieske et al.}

\begin{document}
\submitted{Received October 12, 2005; accepted December 15, 2005.}

\title{Spectroscopic metallicities for Fornax UCD\lowercase{s}, GC\lowercase{s} and \lowercase{d}E,N\lowercase{s}}

\author{S. Mieske\altaffilmark{1},
M. Hilker\altaffilmark{2}, 
L. Infante\altaffilmark{3} and 
A. Jord\'an\altaffilmark{1,4}}
\altaffiltext{1}{European Southern Observatory, Karl-Schwarzschild-Strasse 2, 85748 Garching bei M\"unchen, Germany; smieske@eso.org, ajordan@eso.org}

\altaffiltext{2}{Sternwarte der Universit\"at Bonn, 
Auf dem H\"ugel 71, 53121 Bonn, Germany; mhilker@astro.uni-bonn.de}

\altaffiltext{3}{Departamento de Astronom\'{\i}a y Astrof\'{\i}sica, Pontificia
Universidad Cat\'olica de Chile, Casilla 306, Santiago 22, Chile; linfante@astro.puc.cl}

\altaffiltext{4}{Astrophysics, Denys Wilkinson Building, University
of Oxford, 1 Keble Road, OX1 3RH, UK}

\begin{abstract}
\noindent Various formation channels for the puzzling ultra-compact dwarf galaxies (UCDs) have been proposed in the last few years. To better judge on some of the competing scenarios, we present spectroscopic [Fe/H] estimates for a sample of 26 compact objects in the central region of the Fornax cluster, covering the magnitude range of UCDs and bright globular clusters ($18<V<21$ or $-13.4<M_V<-10.4$ mag). 
We find a break in the metallicity distribution of compact objects at $M_V\simeq -11$ mag ($\simeq$ 3$\times 10^6$ M$_{\sun}$): for $M_V<-11$ mag the mean metallicity is [Fe/H]$=-$0.62 $\pm$ 0.05 dex, 0.56 $\pm$ 0.15 dex higher than the value of $-$1.18 $\pm$ 0.15 dex found for $M_V>-11$ mag. This metallicity break is accompanied by a change in the size-luminosity relation for compact objects, as deduced from HST-imaging: for $M_V<-11$ mag, $r_h$ scales with luminosity, while for $M_V>-11$ mag, $r_h$ is almost luminosity-independent. In our study we therefore assume a limiting absolute magnitude of $M_V=-11$ mag between UCDs and globular clusters. 
The mean metallicity of five Fornax dE,N nuclei included in our study is about 0.8 dex lower than that of the UCDs, a difference significant at the 4.5$\sigma$ level. This difference is marginally higher than expected from a comparison of their  $(V-I)$ colors, indicating that UCDs are younger than or at most coeval to dE,N nuclei. Because of the large metallicity discrepancy between UCDs and nuclei we disfavor the hypothesis that most of the Fornax UCDs are the remnant nuclei of tidally stripped dE,Ns. Our metallicity estimates for UCDs are closer to but slightly below those derived for young massive clusters (YMCs) of comparable masses. 
We therefore favor a scenario where most UCDs in Fornax are successors of merged YMCs produced in the course of violent galaxy-galaxy mergers. It is noted that in contrast to that, the properties of Virgo UCDs are more consistent with the stripping scenario, suggesting that {\it different} UCD formation channels may dominate in either cluster.
\end{abstract}

\keywords{galaxies: clusters: individual: Fornax -- galaxies:
dwarf -- galaxies: fundamental parameters -- galaxies: nuclei -- 
globular clusters: general}

\section{Introduction}
\subsection{Discovery of Ultra-compact Dwarf Galaxies}
\label{intro}
\noindent In their spectroscopic studies of the Fornax region, Hilker et al. \cite{Hilker99} and later Drinkwater et
 al. (\cite{Drinkw00} and \cite{Drinkw03}) reported
 on the discovery of six isolated compact stellar systems in the Fornax cluster,
having $-13.4<M_V<-12$ mag and half-light radii $r_h$ between 15 and 30 pc. These objects had been interpreted as galactic stars in previous imaging surveys because of their star-like appearance on photographic plates. Due to their compactness at luminosities comparable to average dwarf galaxies, they were dubbed ``ultra-compact dwarf galaxies`` (UCDs) (Phillips et al.~\cite{Philli01}). High resolution imaging and spectroscopy (Drinkwater et al.~\cite{Drinkw03}) place the Fornax UCDs between the sequence of globular clusters (GCs) and
dwarf elliptical galaxies (dEs) in the fundamental plane of stellar 
systems. Their M/L ratios are in the range 2-4, at the upper end of values obtained for globular clusters. The luminosities and sizes of UCDs are comparable to those of 
bright nuclei of nucleated dwarf ellipticals (dE,Ns) (Lotz et al.~\cite{Lotz04}).

It is not yet clear how objects as compact and luminous as the UCDs may have formed. One much 
discussed hypothesis is the stripping scenario (Bekki et al.~\cite{Bekki03}), which proposes that UCDs are remnant nuclei of dwarf galaxies stripped in the potential well of their host cluster. It has also been proposed that most UCDs are simply very bright globular clusters (Mieske et al.~\cite{Mieske02}), although we note that the sizes of the brightest Fornax UCDs are much larger than those of globular clusters (Drinkwater et al.~\cite{Drinkw03}, Jord\'an et al.~\cite{Jordan05a}). UCDs could also be explained by stellar super-clusters created in gas-rich galaxy galaxy mergers (Fellhauer \& Kroupa~\cite{Fellha02} and~\cite{Fellha05}).
Related to the latter scenario, Bastian et al.~\cite{Bastia05a} propose that young massive clusters (YMCs) created in recent galaxy mergers may resemble UCD properties after a Hubble time. Maraston et al.~\cite{Marast04} show that the young massive cluster W3 in the merger remnant galaxy NGC 7252 will resemble UCD fundamental plane properties if aged to several Gyrs.

Recently, UCDs have also been detected in other environments: UCD candidates brighter than those in Fornax have been found in the massive galaxy
cluster Abell 1689 (Mieske et al.~\cite{Mieske05a}), consistent with the idea that tidal forces play a major role in the creation of UCDs.
Hasegan et al. \cite{Hasega05} report on the discovery of several UCDs in the Virgo cluster, some of which have high M/L ratios between 6 and 9. Those authors suggest a high M/L ratio as a criterion to tell UCDs from globular clusters. Also Jones et al.~\cite{Jones05} have discovered 9 UCD candidates in Virgo, in a survey analogous to the one performed by Drinkwater et al.~\cite{Drinkw00} in Fornax. Like the investigation of Hasegan et al., they find that UCDs in Virgo are close to the blue peak of the GC color distribution, in contrast to Fornax where they coincide with the red peak (Mieske et al.~\cite{Mieske04a}). The data of Jones et al.~\cite{Jones05} give a $(b_j - r)$ colour difference of 0.33 $\pm$ 0.08 mag between Fornax and Virgo UCDs, significant at the 4$\sigma$ level. This color difference hints at possible different formation scenarios, which will be discussed in the course of this paper. De Propris et al.~\cite{deprop05} show from HST imaging that the surface brightness profile of five Fornax UCDs is only matched by the brightest dE,N nuclei in Virgo, while the UCDs are significantly more extended than fainter nuclei. That is not necessarily inconsistent with the stripping scenario, because UCDs have luminosities comparable to the brightest nuclei. Fainter UCDs might have different scale sizes. Furthermore, stripping processes can alter the structural parameters of embedded nuclei (Bekki et al.~\cite{Bekki03}). Therefore, a direct comparison of surface brightness profiles might be a biased way of judging on differences between UCDs and embedded nuclei.

A note regarding nomenclature: the term ``ultra-compact dwarf galaxies'' (Phillipps et al.~\cite{Philli01}) should be understood as a purely {\it morphological} description, referring to astronomical objects of luminosities comparable to average dwarf galaxies and sizes much smaller than those. This term does not imply a specific UCD {\it formation scenario}, be it as remnant of more extended dwarf galaxies or as merged super-clusters. Furthermore, in order to call a source UCD, it should clearly be distinguishable from the bulk of ``normal'' GCs, based on properties like mass, size or stellar populations. In our studies, whenever a separation between GCs and UCDs is not clear, we therefore use the term ``compact objects''.
\subsection{The Fornax Compact Object Survey}
\noindent In the Fornax Compact Objects Survey FCOS (Mieske et al.~\cite{Mieske02}~and~\cite{Mieske04a}) we have analyzed the photometric and kinematic properties of compact objects in the Fornax cluster. Our sample of compact objects was defined via radial velocity cluster membership, spanning a range of $18<V<21$ ($-13.4<M_V<-10.4$) mag. At the bright end, these are the UCDs, while at the faint end the sample is dominated by GCs.

 A main finding of the FCOS was that compact objects with $V<20$ mag have somewhat different kinematics, spatial distribution and photometric properties than compact objects with $V>20$ mag. Furthermore, the latter sample was more consistent with the GC system of NGC 1399 than the former one. This was interpreted such that UCDs dominate the sample for $V<20$ mag, while GCs dominate for $V>20$ mag. Specifically, UCDs (i.e. compact objects with $V<20$ mag) are found to be redder by about 0.10 mag than the overall GC population of NGC 1399. Their colors are shifted about 
0.2 mag red-wards of the well known color magnitude relation for dEs (Hilker et al.~\cite{Hilker03}, Karick et al.~\cite{Karick03}) and consistent with a color magnitude trend of comparable slope. The nuclei of present-day dE,Ns are bluer than UCDs by ~0.10 to 0.15 mag (Lotz et al.~\cite{Lotz04}), suggesting that UCDs either have higher metallicities or older 
integrated stellar populations than the present-day nuclei. 
The magnitude distribution of compact objects observed in the FCOS shows a soft
transition between UCDs and GCs. There is a 2$\sigma$ overpopulation with respect to 
the extrapolated bright end of NGC 1399's GC luminosity function.

The findings obtained from the FCOS are consistent both with the stripping scenario (Bekki et al.~\cite{Bekki03}) and the stellar super-cluster scenario (Fellhauer \& Kroupa~\cite{Fellha02}) as a source of UCDs.
The major uncertainty of the data interpretation for the FCOS is the origin of the color difference between Fornax UCDs, the nuclei of dE,Ns and the main body of GCs. The redder color of UCDs as compared to nuclei either means older integrated ages or higher metallicities. Within the scenario of Bekki et al.~\cite{Bekki03}, the M/L ratio of the stripping remnant (nucleus + remainder of envelope) is lower by a factor of two compared to the progenitor M/L, while the nucleus itself loses about 20\% of its mass. This mass loss and the associated weakening of the gravitational potential will lead to expulsion of a substantial gas fraction out of the stripping remnant, a process conceivably supported by ram-pressure stripping. This would lower the efficiency of or completely halt subsequent star formation events in the naked nucleus. Therefore one would expect the UCDs to have older integrated ages and lower or at most equal metallicity as compared to dE,N nuclei (see also Jones et al.~\cite{Jones05}). One would {\it not} expect in the stripping scenario that the UCDs are on average more metal-rich than still existing dE,N nuclei.
\subsection{Aim of this paper}
\noindent The aim of this paper is to lift the age-metallicity degeneracy inherent in the broad-band color measurements for Fornax UCDs, GCs and dE,N nuclei, in order to better assess the validity of the various UCD formation scenarios. To this end we present new spectroscopic data for compact objects in Fornax, obtained with the 6.5m Magellan telescope at Las Campanas Observatory, Chile. The paper is structured as follows: in Sect.~\ref{data} we present the data and their reduction. Sect.~\ref{cmrel} re-assesses the color-magnitude relation for Fornax compact objects proposed in Mieske et al.~\cite{Mieske04a}.  In Sect.~\ref{results} we present the results of the spectroscopic metallicity measurements for UCDs, GCs and dE,N nuclei. The results are discussed in Sect.~\ref{discussion}. We finish this paper with the conclusions in Sect.~\ref{conclusions}.
\section{The data}
\noindent The data we present in this paper were obtained in the night of 5th to 6th December 2004 at Las Campanas Observatory (LCO), Chile. The instrument used was the Inamori
Magellan Areal Camera and Spectrograph ``IMACS'' in spectroscopy mode with the
``short'' f/2 camera, mounted at the 6.5m Baade telescope. 
The double-asphere, glass-and-oil-lens f/2 camera
produces an image 
of 27.4$'$ field diameter at 0.20 arcsec per pixel. The IMACS detector is an 8192$\times$8192
CCD mosaic camera which uses 8 thinned 2k$\times$4k detectors. 
Gaps of about 50 pixels separate the chips. We used the intermediate resolution 300 grism, which has a dispersion of 1.34 {\AA} per 0.2$''$ pixel and covers a wavelength range of 3900 {\AA} - 10000 {\AA}. At a typical seeing of 0.8$''$, the instrumental resolution was of the order of 5 {\AA}.

\begin{figure}
\plotone{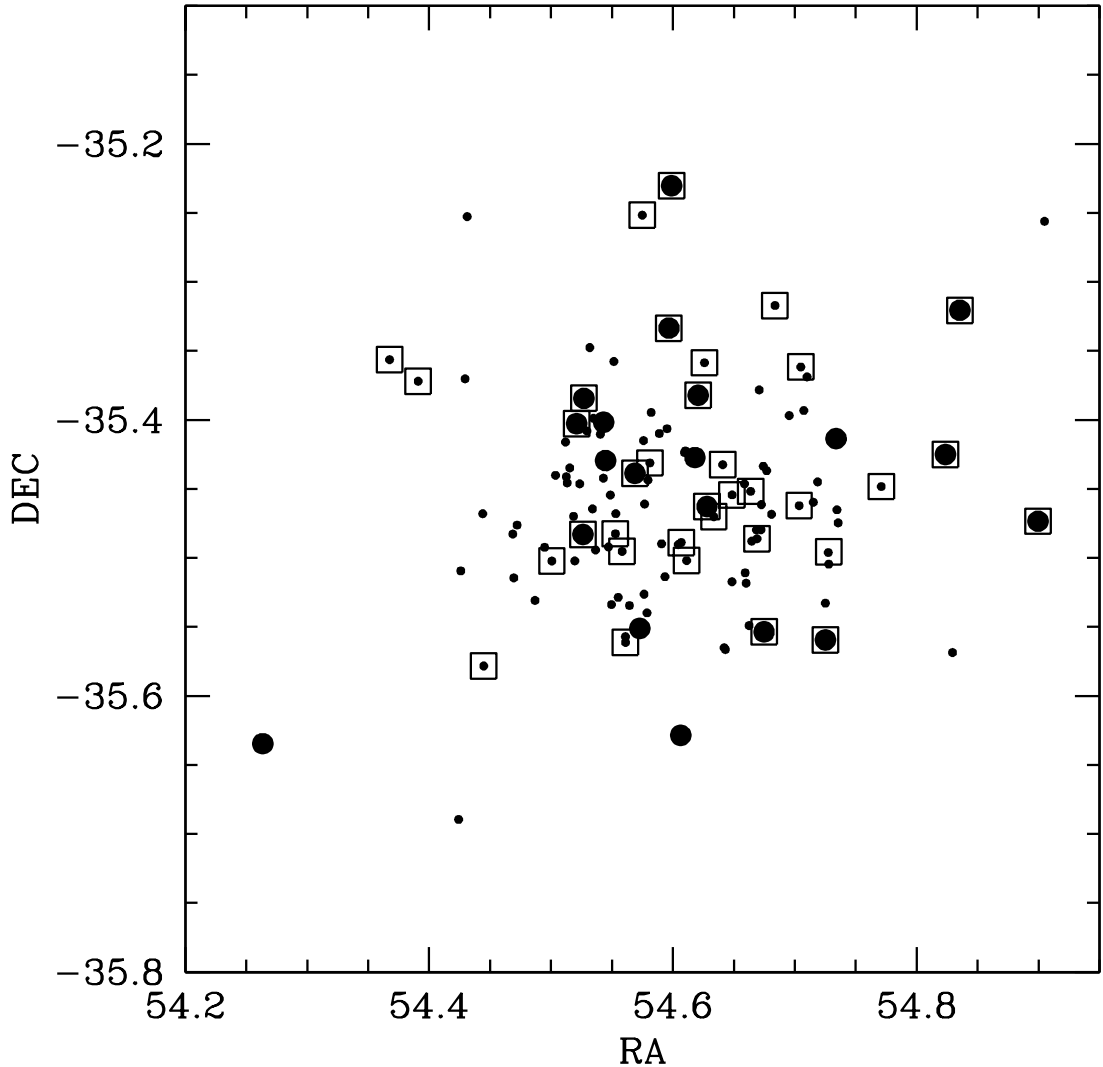}
\caption{\label{map} Map of 118 spectroscopically confirmed compact sources in the central Fornax cluster with $V<22$ mag. The spectroscopic cluster memberships are from Mieske et al.~\cite{Mieske04a} and from Dirsch et al.~\cite{Dirsch04}. Small dots are compact objects fainter than $V=20.4$ mag (see Sect.~\ref{results}). Large dots are compact objects brighter than $V=20.4$ mag. The 35 objects observed with IMACS in this paper are marked by open squares.}
\end{figure}

We observed one MOS mask centered on NGC 1399, targeting 35 compact Fornax members in the magnitude range $18<V<21$ ($-13.4<M_V<-10.4$) mag. Those members were chosen from the FCOS and the investigation of NGC 1399's GCS by Richtler et al.~\cite{Richtl04}. The total integration time was 2.5 hours, split up into three exposures of 30 minutes and three exposures of 20 minutes. See Figs.~\ref{map} and~\ref{cmdall1} for a map and color-magnitude diagram of all compact Fornax members in the magnitude range $18<V<22$ mag. The 35 sources observed with IMACS are especially indicated. The UCDs originally discovered by Drinkwater et al.~\cite{Drinkw00} are those brighter than $V=19.5$ mag. 
\label{data}
\subsection{Data reduction}
\noindent The reduction procedure from raw images to two-dimensional, sky subtracted, wavelength calibrated spectra was performed with the IMACS version of the COSMOS-package\footnote{Carnegie Observatories System for
 MultiObject Spectroscopy, http://llama.lco.cl/aoemler/COSMOS.html}. COSMOS performs the following reduction steps, based on the observation definition file used to generate the slit mask: construct spectral maps, use a comparison arc to adjust spectral maps,  calculate positions of spectral lines using a spectral map, construct a spectroscopic flat field frame, do bias and flat field corrections of frames, subtract sky from spectral frame, extract 2D-spectra, co-add spectral exposures, using a cosmic-ray rejection algorithm.

The extraction of the one-dimensional spectrum from the two-dimensional output of COSMOS was performed with the APALL task within the IRAF TWODSPEC package. To test the absolute accuracy of the wavelength calibration, these 1D-spectra were cross-correlated against the same template spectra as used for the radial velocity measurement in the FCOS. For the 35 compact objects observed with IMACS, the mean radial velocity was insignificantly higher by only 13 $\pm$ 20 km/s than measured in the FCOS.

In order to properly derive Lick line indices, we slightly downgraded our spectra to the Lick system resolution of 9{\AA}. For the brightest object observed, namely UCD 3 with $V=18.03$ mag, the S/N at the peak of the spectral throughput was around 100.
We derived Lick indices for our 35 observed sources using the latest Lick pass-band definitions of Trager et al.~\cite{Trager98}, appropriately red-shifted according to the radial velocity of each investigated object.

For converting the line index measurements to [Fe/H], we used the empirical calibration of Puzia et al.~\cite{Puzia02} between [Fe/H] and several metal line indices for a large sample of metal-poor and metal-rich Galactic globular clusters, based on the Zinn \& West~\cite{Zinn84} scale. Specifically, we employed  the four most strongly correlated indices Mg2, Mgb, $<Fe>$, $[MgFe]$. For each of those indices we derived [Fe/H] according to Puzia's calibration relation and we attributed a weight corresponding to the inverse square of the rms of the calibration relation. Then, we adopted the weighted mean of the four measurements as [Fe/H]. The scatter of the four values around their mean is adopted as error estimate for [Fe/H].

The results of these measurements are given in Table~\ref{FeHtable} and plotted in Figs.~\ref{Fe_V_nucl} and~\ref{VI_Fe_V}. Fig.~\ref{Hb_Mg2} shows the age indicator H$\beta$ vs. [Fe/H]. See Sect.~\ref{results} for further details.
\section{Re-assessment of color-magnitude relation}
\label{cmrel}
\noindent Before proceeding to the metallicity distribution of compact objects in Fornax, we revisit the possibility of a color-magnitude relation as suggested by Mieske et al.~\cite{Mieske04a}.

\begin{figure}
\plotone{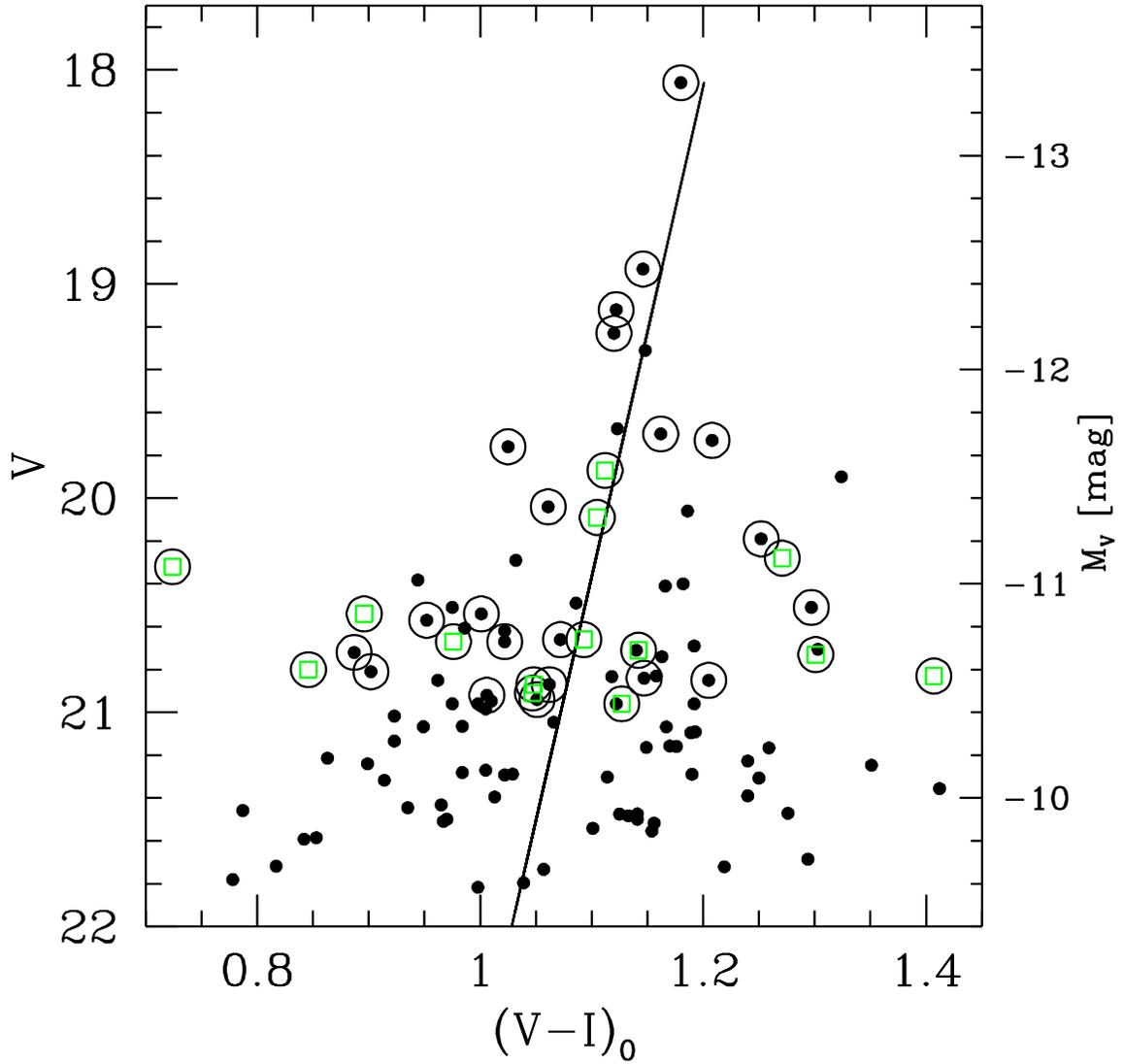}
\caption{\label{cmdall1}Dots show the color-magnitude distribution of those 101 sources from Fig.~\ref{map} with $VI$ photometry. The black line is a linear fit to these points (slope significant at 2.5$\sigma$). Open squares indicate objects with known cluster membership for which we do not have $VI$ photometry. Instead, the Washington $CR$ photometry values of Dirsch et al.~\cite{Dirsch04} are shown, transformed to the $VI$ bands (see Mieske et al.~\cite{Mieske04a}). The 35 objects observed with IMACS in this paper are marked by circles.}
\end{figure}

Fig.~\ref{cmdall1} shows a $VI$ CMD of compact Fornax members with $V<22$ ($M_V<-9.5$) mag, i.e. about two magnitudes brighter than the turn-over of the globular cluster luminosity function (GCLF). The difference to the CMD shown in Mieske et al.~\cite{Mieske04a} is that now we include additional confirmed cluster members from Richtler et al.~\cite{Richtl04} and Dirsch et al.~\cite{Dirsch04} in the magnitude range $19.7<V<22$ mag. Photometry is from Hilker et al.~\cite{Hilker03} for all sources. Over-plotted is a linear fit to all data points, applying a 2.5$\sigma$  rejection algorithm. For this fitting, the IRAF task NFIT1d within the STSDAS package was used. We have slightly modified the error computation for the fitted coefficients as compared to Mieske et al.~\cite{Mieske04a}. In the latter paper the coefficient errors were calculated in the following way: after fitting the function, the independent variable vector $(V-I)$ is replicated 15 times, each data point being replaced by the fitted function value plus noise with dispersion given by {\it the error bar of each data point} (as determined by artificial star experiments in Mieske et al.~\cite{Mieske04a}). This does not take into account the fact that there are {\it physical} variations in color that may go beyond the level of the photometric accuracy. A more conservative error estimate is achieved when the dispersion for the re-sampling is adopted as the {\it dispersion of the original data around the fit.} The significances quoted in the following are based on this error estimate.

As can be seen in Fig.~\ref{cmdall1}, for the sample as a whole there is a trend of redder color with increasing luminosity, significant at the 2.5$\sigma$ level (compared to 3.0$\sigma$ in Mieske et al.~\cite{Mieske04a}). The significance of this trend remains unchanged when rejecting the brightest data point from the sample. However, the slope becomes insignificant when restricting the sample to the sources with $V<20$ mag, due to low number statistics.
Interestingly, the slope remains almost unchanged at even slightly higher significance (2.8$\sigma$) when excluding all sources with $V<20$ mag. The difference in color between the faint ($V>20$ mag) and bright ($V<20$ mag) subsample is 0.080 $\pm$ 0.024 mag, significant at the 3.4$\sigma$ level. All this is consistent with a gradual color-magnitude trend over the entire investigated magnitude range $18<V<22$ mag. However, the significance levels are still not very high, and spectroscopic information is indispensable to clarify whether the color shift is due to age or metallicity.
\section{Results}
\label{results}
\begin{figure}
\epsscale{.8}
\plotone{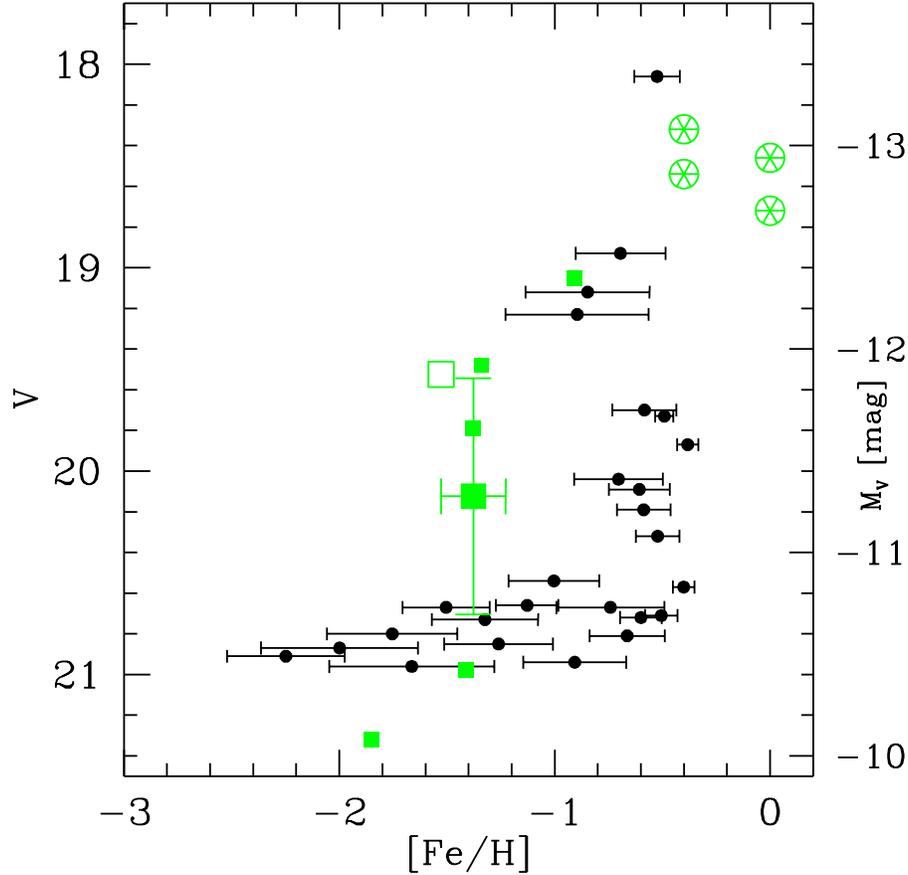}
\epsscale{1}
\caption{\label{Fe_V_nucl}[Fe/H] in dex plotted vs. $V$ magnitude for 26 out of the 35 sources re-observed with IMACS (see Fig.~\ref{cmdall1}). The nine objects not plotted have too low S/N for reliable line index measurement. Note the break in the metallicity distribution at about $M_V=-11$ mag. The small filled squares give [Fe/H] and estimated nuclear magnitudes for 5 Fornax cluster dE,Ns (two [Fe/H] measurements from this paper, three from Mieske et al.~\cite{Mieske02}). Nuclear magnitudes could not be accurately determined from our low-resolution photometry. Instead, we plot the galaxy magnitude + a mean offset of 4.14 mag between galaxy and nucleus, as derived in the HST-study by Lotz et al.~\cite{Lotz04} for a sample of Fornax/Virgo dE,Ns disjunct with our sample. The large filled square gives the mean of all 5 nuclear regions.
The open square indicates the mean value for the five dE,Ns if corrected for observational biases and extrapolated by 0.6 mag to the mean UCD magnitude (see Sect.~\ref{compnuclei} for further details). The encircled asterisks indicate the location of the young massive cluster W3 in NGC 7252 if aged passively to 10 and 13 Gyrs, with assumed metallicities Z=0.4Z$\sun$ and Z=1.0Z$\sun$ (Hilker et al. in preparation, using Bruzual \& Charlot~\cite{Bruzua03}). Starting points are its present observed absolute magnitude $M_V\simeq -$16.2 mag and an assumed age of 300 Myr (Maraston et al.~\cite{Marast04}).}
\end{figure}

\begin{figure}
\plotone{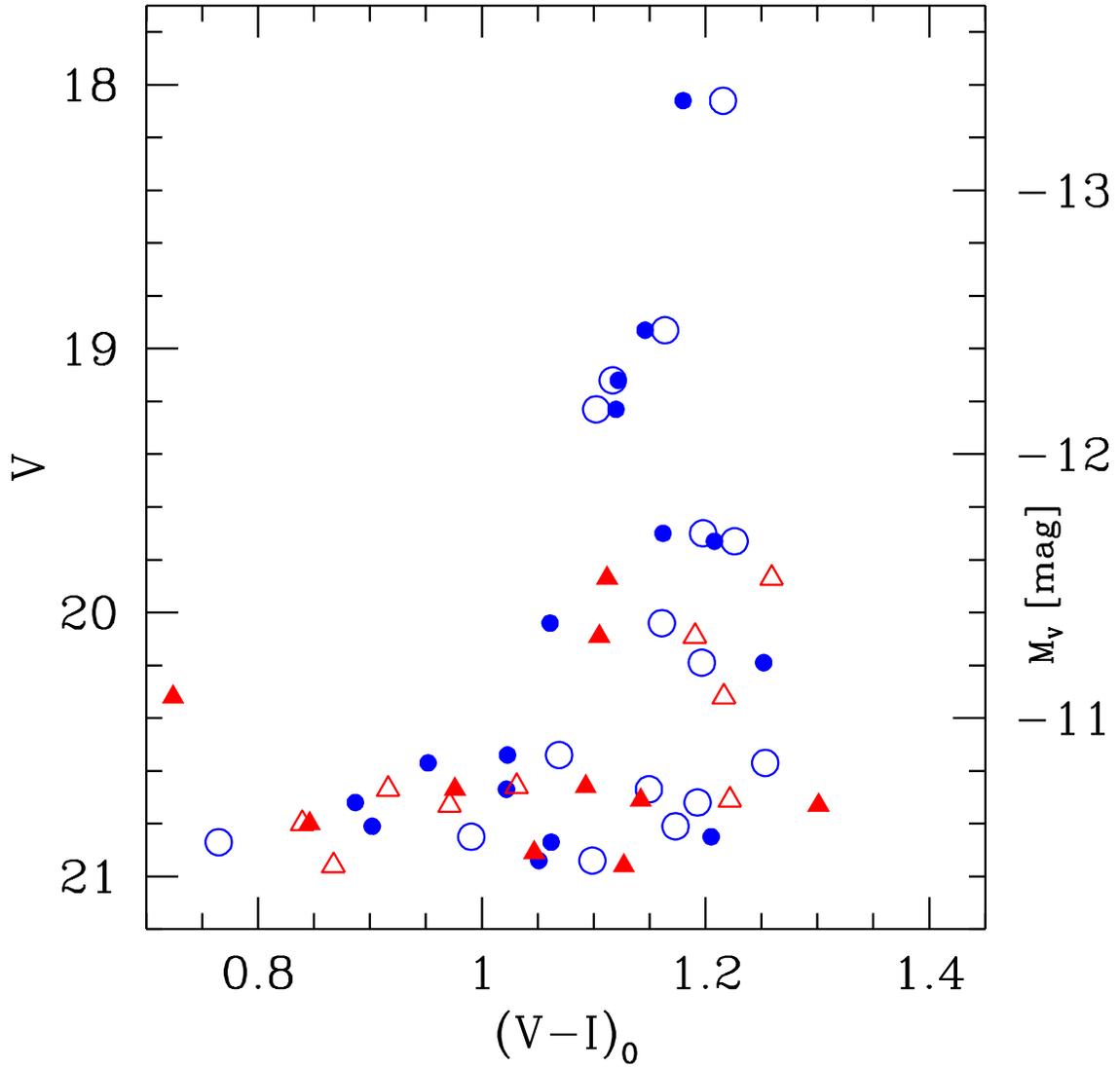}
\caption{\label{VI_Fe_V}Dots: CMD of the sources in Fig.~\ref{Fe_V_nucl} with $VI$ photometry available. Circles: the measured [Fe/H] converted to $(V-I)$ using the [Fe/H]$-(V-I)$ relation by Kissler-Patig et al.~\cite{Kissle98} (valid for [Fe/H]$>-1.2$ dex). Note the very good agreement for the six brightest sources. Filled triangles: CMD of those sources for which only $CR$ photometry is available, tranformed to $VI$ according to Mieske et al.~\cite{Mieske04a}. Open triangles: the measured [Fe/H] converted to $(V-I)$ using the [Fe/H]$-(V-I)$ relation by Kissler-Patig et al.~\cite{Kissle98}}
\end{figure}

\noindent In Fig.~\ref{Fe_V_nucl} we plot [Fe/H] vs. apparent $V$ magnitude for 26 out of the 35 observed sources. The 9 objects not plotted had too low S/N for line index measurements. Over-plotted are values for 5 nuclei of dE,Ns in the Fornax cluster. Two of those measurements are from the data set in this paper, three from the first part of FCOS (Mieske et al.~\cite{Mieske02}).

One striking feature of Fig.~\ref{Fe_V_nucl} is a metallicity break at $M_V\simeq -11$ mag: among the brightest compact objects in Fornax there are no metal-poor ones. When separating the sample into bright and faint with a limit of $V=20$ mag -- inspired by the results of the FCOS --, the mean [Fe/H] of the bright sample is $-$0.63 $\pm$ 0.07 dex, that of the faint sample $-$1.14 $\pm$ 0.15 dex. The difference then is 0.51 $\pm$ 0.16 dex, significant at the 3.2$\sigma$ level. If instead of $V=20$ mag adopting $V=20.4$ mag ($M_V=-11$ mag) as the limit, the values for the bright and faint sub-samples are $-$0.62 $\pm$ 0.05 dex and $-$1.18 $\pm$ 0.15 dex, respectively. The difference rises to 0.56 $\pm$ 0.15 dex, somewhat more significant at the 3.7$\sigma$ level. These confidence levels are of the same order than the difference in {\it color} $(V-I)$ between bright and faint subsamples. Note that while we can trace back the color {\it shift} to a metallicity {\it shift}, the size of our IMACS sample is not large enough to judge on the existence of a {\it continuous} luminosity-metallicity relation.

To get some feeling for possible systematic biases in our derivations of [Fe/H] -- i.e. offsets of our line index measurements with respect to the Lick system -- consider the CMD in Fig.~\ref{VI_Fe_V}. Here, we show the {\it measured} $(V-I)$ values plus the measured [Fe/H] values {\it transformed} to $(V-I)$, using the transformation relation by Kissler-Patig et al.~\cite{Kissle98}, which was calculated for GCs in NGC 1399 with metallicities above [Fe/H]$\simeq-1.2$ dex. For the eight sources brighter than $M_V=-11$ mag with $VI$ photometry available (all of which have [Fe/H]$>-1$ dex) there is an offset between measured and transformed $(V-I)$ of 0.018 $\pm$ 0.016 mag with an rms scatter of 0.045 mag. Under the assumption that the Lick calibration itself between line indices and [Fe/H] is correct, this indicates an upper limit to a potential systematic bias in [Fe/H] of about 0.10 dex.

The next question to ask is: does the color difference in $(V-I)$ between bright and faint sub-sample ($M_V=-11$ mag as limit) correspond to the measured metallicity-difference? The mean color for the bright sub-sample, taking only objects with $VI$ photometry available, is 1.15 $\pm$ 0.02 mag. For the faint sub-sample it is 1.01 $\pm$ 0.04 mag. To convert this difference of $\Delta (V-I)=0.14 \pm$ 0.04 mag into an expected metallicity difference (under the assumption of two coeval populations), we again apply the transformation of Kissler-Patig et al.~\cite{Kissle98}, specifically their calibration slope $\frac{d[Fe/H]}{d(V-I)}= $3.27 $\pm$ 0.32. We would then expect a [Fe/H] difference of 0.46 $\pm$ 0.15 dex, in very good agreement with the actually found difference of 0.56 $\pm$ 0.15 dex. Note that the mean metallicity of the faint sub-sample is at about the limit of the linear approximation by Kissler-Patig et al.~\cite{Kissle98}. For lower metallicities, the slope becomes steeper. Kissler-Patig et al.~\cite{Kissle97} found a slope twice as high for the metal-poor Milky Way globular clusters (see also Couture et al.~\cite{Coutur90}, Kundu \& Whitmore~\cite{Kundu98}). Taking this steepening of the slope into account, the expected metallicity difference is higher by about 0.1 to 0.2 dex, bringing it to an even better match with the observations. 
Thus, we can state that our data are consistent with the assumption that the only difference between bright and faint sub-sample is metallicity.

\begin{figure}
\plotone{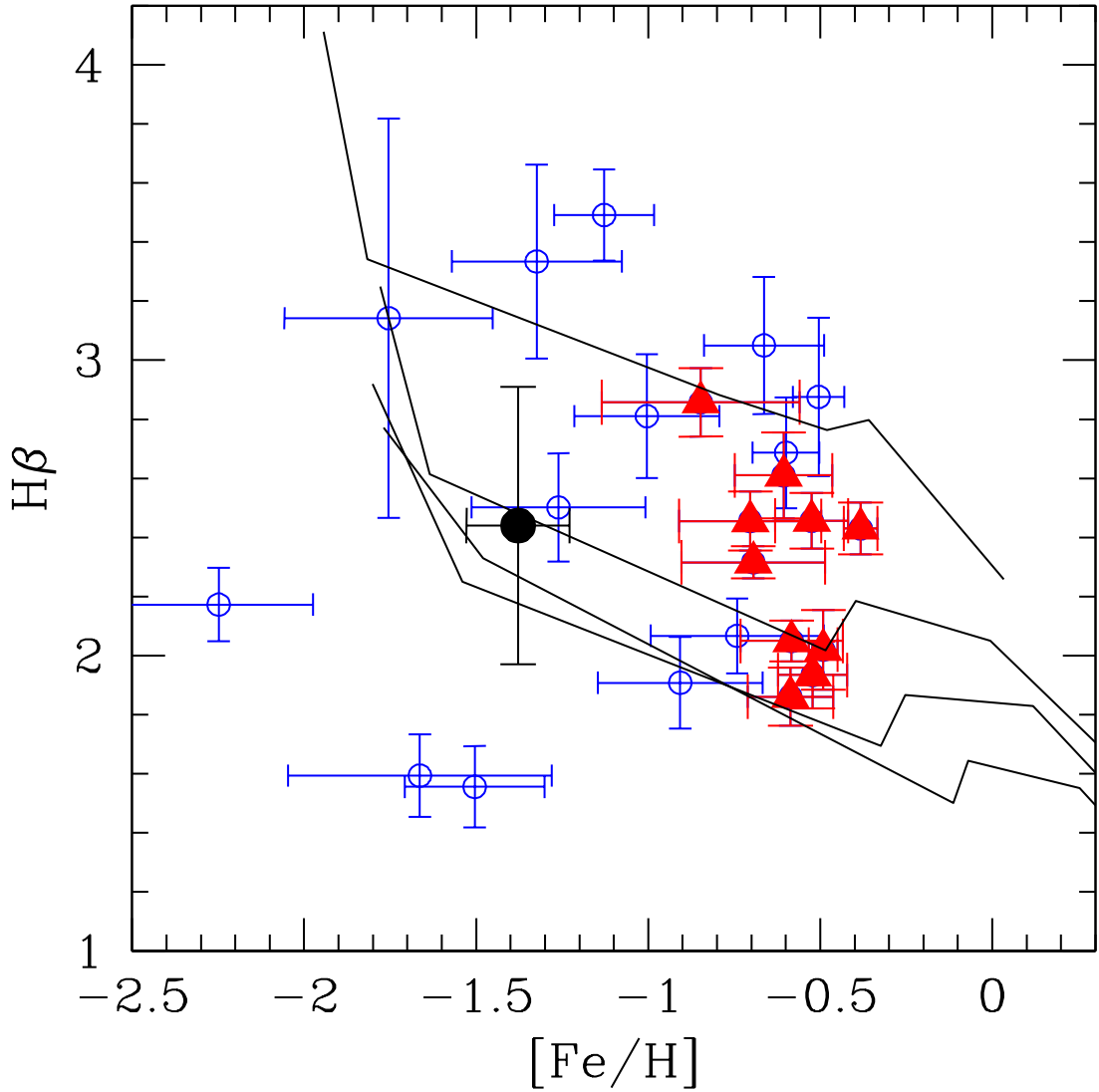}
\caption{\label{Hb_Mg2}H$\beta$ equivalent width plotted vs. [Fe/H] for the sources from Fig.~\ref{Fe_V_nucl}. Three of the sources from Fig.~\ref{Fe_V_nucl} do not have a H$\beta$ measurement due to chip gaps. Sources marked by filled triangles are those with $V<20.4$ mag. The large filled circle indicates the mean for the five dE,Ns shown in Fig.~\ref{Fe_V_nucl}. Overplotted for comparison are model grids from Bruzual \& Charlot (\cite{Bruzua03}) for single age stellar populations, covering ages between 2 and 13 Gyrs (from top to bottom) and [Fe/H] from -2.25 to 0.56 dex (from left to right).}
\end{figure}

Are there also significant differences in age sensitive line indices between the bright and faint sub-sample (Fig.~\ref{Hb_Mg2})? The mean $H\beta$ equivalent width for sources with $V<20.4$ mag is 2.30 $\pm$ 0.10 {\AA} with a dispersion of 0.3 {\AA}, compared to 2.69 $\pm$ 0.20 {\AA} and a dispersion of 0.6 {\AA} for $V>20.4$ mag, Mg2$>0$ and $H\beta >1$. For a proper age comparison, one also has to take into account the slope of isochrones in the [Fe/H]$-H\beta$ plane, which is approximately $\frac{dH\beta}{d[Fe/H]}=-0.5$ (see Fig.~\ref{Hb_Mg2}). With a metallicity difference of 0.55 dex as estimated above, one would expect a difference in $H\beta$ of 0.25 to 0.3 {\AA} between the bright and faint sample in the case of coeval populations. The actual difference is 0.39 $\pm$ 0.22 {\AA}. The data are therefore consistent with coeval populations, but also allow for age differences of a few Gyrs in both directions. The UCDs appear most consistent with intermediate age populations.

Concluding this section: we have found a $\simeq$ 0.55 dex higher metallicity for compact objects in Fornax with $M_V<-11$ mag than for those with $M_V>-11$ mag, significant at the 3.7$\sigma$ level and in good agreement with their color difference. Our findings hence support the hypothesis that for $M_V<-11$ mag, the population of compact objects has a different formation history than the main body of NGC 1399's GCS.
In the following, we will refer as UCDs to compact objects with $M_V<-11$ mag and as GCs to those with $M_V>-11$ mag. With respect to this separation it is worth noting that the metallicity
range of GCs includes that of
the UCDs, and therefore we cannot exclude that a low-luminosity extension of the UCDs somewhat contaminates the GC sample. In Sect.~\ref{discussion} the separation at $M_V=-11$ mag will be confirmed by a size comparison from HST imaging, indicating that UCDs are probably not a {\it major} fraction of Fornax compact objects with $M_V>-11$ mag.
\subsection{Comparison with dE,N nuclei}
\label{compnuclei}
\noindent One much discussed formation scenario for the UCDs is that they are the compact remnants of tidally stripped dE,Ns (e.g. Bekki et al.~\cite{Bekki03}, Drinkwater et al.~\cite{Drinkw03}, see also Bassino et al.~\cite{Bassin94}). 
Fig.~\ref{Fe_V_nucl} compares the metallicities of bright GCs plus UCDs with those of the central regions of five Fornax cluster dE,Ns. The central regions of the investigated dE,Ns have a mean metallicity of [Fe/H]$=-$1.38 $\pm$ 0.15 dex, 0.75 $\pm$ 0.17 dex lower than the mean UCD metallicity. This is a 4.5$\sigma$ difference. Literature estimates of [Fe/H] for Fornax dEs and dE,Ns (Held \& Mould~\cite{Held94}, Brodie \& Huchra~\cite{Brodie91}, Rakos et al. \cite{Rakos01}) also show lower average metallicities than for the UCDs, but are about 0.1 to 0.2 mag higher than our values (see also Mieske et al.~\cite{Mieske02}). Before proceeding further, we draw the attention to two effects that potentially bias our result:

1. In addition to the light from the nucleus itself, a certain fraction of the stellar envelope of the investigated dE,Ns also contributes to the measured signal. That fraction was on average about 40\% for the five dE,Ns investigated (scattering between 10\% and 85\%), as judged from the slit size and width (this paper), total galaxy luminosity (Hilker et al.~\cite{Hilker03}) and nuclear luminosity (this paper and Lotz et al.~\cite{Lotz04}). This contribution becomes important in the case of significant differences between the stellar populations of the dE,N envelope and the nucleus. Lotz et al.~\cite{Lotz04} have shown that on average, dE,N envelopes are redder by 0.1 to 0.2 mag than their nuclei. If this is due to a metallicity difference (as argued by Lotz et al.), our measurements will overestimate the nuclear metallicity. To estimate the extent of this overestimation, we assume the mean of the two different slopes derived in the metal-poor ([Fe/H]$<-1.2$) and metal-rich regimes by Kissler-Patig et al.~\cite{Kissle97} and~\cite{Kissle98}, which is $\frac{d[Fe/H]}{d(V-I)}~\simeq$5. The over-estimation of the nuclear metallicity is then about 0.3 dex. This effect can explain why we obtain a slightly lower dE,N metallicity than previous estimates: due to the better spatial resolution of our data, the contribution of the presumably more metal-rich stellar envelope is lower than in the data of the authors in the previous paragraph.

2. Our sample of dE,N nuclei is biased towards lower luminosities compared to the UCDs: the average luminosity of the five nuclei plotted in Fig.~\ref{Fe_V_nucl} is V=20.2 mag, while the UCDs have a mean apparent magnitude of V=19.6. That magnitude difference is relevant because of the color-magnitude relation for nuclei (Lotz et al.~\cite{Lotz04}), which is commonly interpreted as the representation of a mass-metallicity relation. Due to this relation, our sample of five lower luminosity nuclei statistically exhibits a lower metallicity than nuclei of UCD luminosity. With a slope of $\frac{d(V-I)}{dV}\simeq -$0.04 (Lotz et al.~\cite{Lotz04}) and $\frac{d[Fe/H]}{d(V-I)}\simeq$5 (Kissler-Patig et al.~\cite{Kissle97} and \cite{Kissle98}), one would expect a higher metallicity by about 0.12 dex for nuclei of UCD luminosity.

The net effect of the two biases is a slight increase of the metallicity difference by about 0.1-0.2 dex. We therefore confirm a metallicity difference between UCDs and nuclei of at least 0.80 dex.

Already in Mieske et al.~\cite{Mieske04a} and Mieske et al.~\cite{Mieske05b} it was noted that dE,N nuclei from the HST photometry of Lotz et al.~\cite{Lotz04} are bluer than the Fornax UCDs by 0.10 to 0.15 mag, consistent with a lower metallicity as found here. Assuming again $\frac{d[Fe/H]}{d(V-I)}~\simeq $5 (Kissler-Patig et al.~\cite{Kissle97} and~\cite{Kissle98}) one would expect a metallicity difference of the order 0.5 to 0.75 dex. This is consistent with, but at the low end of our measured range. See Sect.~\ref{discussion} for further discussion.
We finally note that the [Fe/H]$-H\beta$ plot in Fig.~\ref{Hb_Mg2} shows no indication for an age difference between the five dE,N nuclei and the UCDs. However, the large scatter of data points also allows for substantial age differences of several Gyrs.
\section{Discussion}
\noindent In the light of the new metallicity estimates outlined in the previous section, we discuss three possible UCD formation channels: 1. Unusually bright GCs. 2. Nuclei of dE,Ns. 3. Young massive star clusters created in violent star-forming events.\\

1. Bright GCs: the brightest Fornax UCD (UCD 3) cannot be defined as a GC, given its very high luminosity ($M_V\simeq$ -13.4 mag) and large half-light radius $r_h$ of 30 pc (Drinkwater et al.~\cite{Drinkw03}). Surface brightness profile fitting even revealed a second exponential component of $r_h \simeq$ 60 pc (Drinkwater et al.~\cite{Drinkw03}). Indeed, UCD 3 is the only UCD that is marginally resolved on ground-based images (see Hilker et al.~\cite{Hilker99} and Mieske et al.~\cite{Mieske04a}).
Furthermore, Fornax UCDs are more extended around the cluster center, more numerous than expected from the extrapolation of a Gaussian GCLF and slightly offset in radial velocity with respect to the GCS (Mieske et al.~\cite{Mieske04a}). The significances of these differences are all of the order 2$\sigma$.

The strongest distinguishing feature is the metallicity difference between UCDs ($M_V<-11$) and GCs ($M_V>-11$ mag) found in the present paper, significant at 3.7$\sigma$. Note, however, that this only implies that Fornax UCDs do not belong to the {\it metal-poor}, i.e. blue population of GCs. The mean metallicity of the Fornax UCDs is close to (yet slightly lower than) the typical value found for the metal-rich, red peak of the GC distribution (e.g. Peng et al.~\cite{Peng05}, Kissler-Patig et al.~\cite{Kissle98}) of about -0.6 to -0.3 dex. Also their mean color of $(V-I)=1.15$ $\pm$ 0.02 is identical to the red peak of NGC 1399's bimodal GC color distribution (Gebhardt \& Kissler-Patig (\cite{Gebhar99}). 
That is: the question ``can one explain (some) UCDs as bright GCs?'' must in the Fornax cluster case be specified as: ``can the red 'bulge GCs' extend to much higher masses/luminosities than the blue GCs?''.

A possibility is that the most massive metal-rich star clusters preferentially form by the merging of several single, already massive clusters in the course of violent gas-rich merger events (Fellhauer\& Kroupa~\cite{Fellha02}). This is an attractive scenario if one assumes that indeed galaxy-galaxy mergers were the main creation channel for the entire metal-rich GC bulge population of GCs (e.g. Rhode \& Zepf~\cite{Rhode04}, Li et al.~\cite{Li04}). This is opposed to the idea that both metal-poor and metal-rich GCs formed in single, monolithic collapse-like, subsequent episodes (e.g. Forbes et al.~\cite{Forbes97}, Harris \& Pudritz~\cite{Harris94}). In the latter case one might expect a less violent, more homogeneous star formation, not necessarily leading to the creation of super-massive GCs.

Kravtsov \& Gnedin \cite{Kravts05} find in their simulations on GCS formation that the maximum globular cluster mass in a given region strongly correlates with the local average star formation rate. On the observational side, Larsen~\cite{Larsen02} correspondingly found that the maximum mass of young massive clusters in spiral galaxies scales with local star formation rate. This implies that the most massive GCs are created in the most violent merger events. Kravtsov \& Gnedin also show that the mass of the most massive GC built in a DM halo correlates with this DM halo's mass, although with a large dispersion, due to the varying star formation conditions within the halo. According to formula (8) of their paper, GCs as massive as the UCDs (1-5$\times$10$^7$M$_{\sun}$) must have progenitor haloes of $\simeq$ 5$\times$10$^{11}$M$_{\sun}$. This is only a few percent of NGC 1399's DM halo mass ($M\simeq 10^{13}M_{\sun}$, see Richtler et al.~\cite{Richtl04}). Therefore one may at face value not even require to merge {\it several} massive GCs to end up with {\it one single} UCD.

However, a strong argument against the identity UCDs=bright single GCs comes from a size comparison: the five brightest Fornax UCDs have King half-light radii $r_h$ between 15 and 30 pc (Drinkwater et al.~\cite{Drinkw03}. The typical $r_h$ of globular clusters is only $\sim$3 pc (Jord\'an et al.~\cite{Jordan05a}) over several orders of magnitude in mass, and even the most massive GCs are not expected to exceed radii of about 5-7 pc (Kravtsov \& Gnedin~\cite{Kravts05}).

The size distribution of UCDs vs. GCs deserves some more detailed investigation: in Fig.~\ref{size} we plot $r_h$ vs. luminosity for those compact objects that were imaged in the ACS Fornax Cluster Survey (Jord\'an~\cite{Jordan05b} and Jord\'an et al. in preparation). All of them are imaged in two ACS pointings. We include the data points for the five brightest UCDs from Drinkwater et al.~\cite{Drinkw03} and also plot the data points for all objects detected in the two ACS pointings. Note that the bulk of sources with $z>20$ mag are GCs, having the typical $r_h$ values around 3 pc (Jord\'an et al.~\cite{Jordan05a}, McLaughlin~\cite{McLaug00}). The fact that they have a luminosity independent mean
size is a physical effect, and not due to the resolution limit of the camera. At the distance of Fornax, 3
pc subtend  ~0.03$''$, or ~0.6 WFC pixels. As shown in Jord\'an
et al.~\cite{Jordan05a}, the resolution limit regarding $r_h$ measurements is ~0.25
WFC pixels ($\simeq$1pc at Virgo/Fornax), which is substantially lower (see also Fig.~\ref{size}).

It is very remarkable that the luminosity of the metallicity break -- corresponding to $M_V\simeq -11$ mag -- coincides exactly with the luminosity brighter than which there is a clear up-turn in the size-luminosity distribution of Fornax compact objects. Brighter than $M_V=-11$ mag one enters the regime where size scales with mass/luminosity, while fainter than that, size is independent of mass/luminosity. This strengthens the separation occurring at $M_V=-11$ mag between a) metal-rich objects at brighter magnitudes whose size scales with luminosity -- the UCDs -- and b) more metal-poor objects at fainter luminosities with luminosity independent sizes -- the GCs. This luminosity limit corresponds to about 3$\times 10^6 M_{\sun}$, assuming an average M/L ratio of 3 (Drinkwater et al.~\cite{Drinkw03}). Note that $\omega$ Centauri, the largest star cluster of the Milky Way, has a mass of about this value (Meylan \& Mayor~\cite{Meylan86}, Pryor \& Meylan~\cite{Pryor93}). Furthermore, this is also the mass which Hasegan et al.~\cite{Hasega05} propose to mark the transition between GCs and UCDs.

A possible clue to the up-turn in the size-luminosity distribution are the findings of  Fellhauer \& Kroupa~\cite{Fellha02} and Bekki et al.~\cite{Bekki04}: super-clusters consisting of several single clusters that merged together will be significantly more extended than the average progenitor cluster. That is, they will deviate towards larger sizes from the constant mass-size relation of globular clusters. The corresponding effect in the Fundamental plane (Drinkwater et al.~\cite{Drinkw03}, Kissler-Patig et al.~\cite{Kissle05}, Bastian et al.~\cite{Bastia05c}, Maraston et al.~\cite{Marast04}) has been found for the brightest and most massive young massive clusters and UCDs, which do not follow the steep M-$\sigma$ relation defined by GCs, but rather the extrapolation of the Faber-Jackson relation.

This supports the hypothesis that UCDs are successors of merged super-clusters, as opposed to being single super-massive GCs. The possibility of young massive clusters in the present universe as examples of ``early'' UCDs will be discussed further in point 3.\\

2. Nuclei: UCDs in Fornax are significantly more metal-rich and redder than nuclei of existing dE,Ns, with no indications for a significant age difference (see previous section). If UCDs are stripped nuclei, they were stripped of their gas and parts of their dark matter halo a long time ago (see Bekki et al.~\cite{Bekki03}). Consequently they have had no opportunity of self-enrichment and star formation since the time of their stripping. In comparison to still existing nuclei that are embedded in a deep potential well with the principal possibility of further star formation events, one would therefore expect equal or lower metallicities and higher integrated ages for ``naked'' nuclei (see also Sect.~\ref{intro}, Mieske et al.~\cite{Mieske04a} and Jones et al.~\cite{Jones05}). We find significantly higher metallicities, but cannot tightly constrain the age difference. The large metallicity difference is not consistent with the stripping scenario as the major creation channel for UCDs in Fornax.
We have outlined in Sect.~\ref{compnuclei} that the metallicity difference between nuclei and UCDs is slightly larger than predicted from their color difference of 0.10 to 0.15 mag. If one allows for age differences between UCDs and nuclei to account for this ``deficit'' in color difference, one would require that UCDs be younger than nuclei. This goes into the opposite direction of what would be expected in the stripping scenario. 

One conceivable way of explaining the metallicity difference between surviving ``embedded'' nuclei and UCDs is that tidal stripping selects the most metal-rich dE,Ns. That is not pure speculation, since in some galaxy clusters, galaxy metallicities have been shown to increase towards the cluster center (e.g. Carter et al.~\cite{Carter02} for Coma, Ezawa et al.~\cite{Ezawa97} for AWM 7). This can lead to a selection bias because tidal stripping should affect only those dE,Ns that orbit close to the cluster center (Bekki et al.~\cite{Bekki03}). However, the amplitude of the metallicity gradients found in the cited studies is only of the order of 0.2-0.3 dex over the entire radial cluster range. That is not sufficient to explain the difference of $\simeq$ 0.80 dex between UCDs and nuclei found by us. Furthermore, Lotz et al.~\cite{Lotz04} do not find any radial color gradient for Fornax cluster dwarf elliptical galaxies, making a strong stripping selection bias unlikely.

Note that the UCDs discovered in the Virgo cluster exhibit slightly different properties, more consistent with the stripping scenario: the UCD candidates detected in the Virgo Cluster Survey (C\^{o}t\'{e} et al.~\cite{Cote04}) based on their larger sizes than average GCs (Hasegan et al.~\cite{Hasega05}) are significantly bluer than the Fornax UCDs (see also Jones et al.~\cite{Jones05}). They are almost as blue as the blue peak of M87's GC color distribution (Peng et al.~\cite{Peng05}) and are consistent with the colors of Virgo dE,N nuclei (C\^{o}t\'{e} et al.~\cite{Cote05}). In agreement with this finding, Evstigneeva et al.~\cite{Evstig05} report line index estimates for Virgo UCDs in which they are found to be old and metal {\it poor}. For some of the Virgo UCDs, Hasegan et al.~\cite{Hasega05} report high M/L ratios between 6 and 9, which in case of virial equilibrium requires some component of dark matter. Those authors suggested a high M/L ratio as a bona-fide criterion to tell UCDs from ordinary GCs. Hasegan et al.~\cite{Hasega05} find that the two UCD candidates with highest M/L ratios are the most metal-poor ones, strengthening the need for dark matter. In contrast, the Fornax UCDs have lower M/L ratios between between 2 and 4 (Drinkwater et al.~\cite{Drinkw03}) and higher metallicities. This makes the scenario of UCDs as successors of nuclei of dark matter dominated dE,Ns more likely for Virgo than for Fornax.

The concluding remark we draw from this is: for the Fornax UCDs, the dE,N stripping seems a less important formation channel, while for the Virgo UCDs the stripping scenario may be the preferred channel.\\

3. Young massive clusters (YMCs): motivated by the difficulties of the stripping scenario to explain the Fornax UCD metallicities, we discuss the hypothesis that UCDs are successors of young massive clusters created in violent star-forming events. This hypothesis is closely related to the scenario of Fellhauer \& Kroupa~(\cite{Fellha02} and~\cite{Fellha05}) that UCDs may be merged super-clusters created in such mergers (see also point 1).

One example for a possible future UCD is the YMC  W3 in the star forming galaxy NGC 7252 (Maraston et al.~\cite{Marast04}) with a mass of $\simeq$ 10$^8$ M$_{\sun}$ whose fundamental plane parameters place it in the region of UCDs/M32 when passively aged several Gyrs. Fellhauer \& Kroupa~\cite{Fellha05} show that W3 may have formed through the merging of several dozen massive single star clusters. Bastian et al.~\cite{Bastia05c} and Kissler-Patig et al.~\cite{Kissle05} find that also other YMCs populate the same Fundamental Plane region as the UCDs.

Fig.~\ref{Fe_V_nucl} shows the location of W3 in the luminosity-metallicity diagram if aged passively to 10 and 13 Gyrs, respectively. Starting points are its present observed absolute magnitude $M_V\simeq -$16.2 mag, an assumed age of 300 Myr and assumed metallicities of Z=0.4Z$\sun$ and Z=1.0Z$\sun$ (Hilker et al. in preparation, Maraston et al.~\cite{Marast04}). W3 is located close to the regime of the brightest Fornax cluster UCDs, shifted slightly to higher metallicities. Note that Bastian et al.~\cite{Bastia05b} also found metallicities of Z$\ge$0.5Z$\sun$ for several other young and massive star cluster complexes in the Antenna Galaxies (NGC 4038/4039). That is an indication that some of the Fornax UCDs may have been created in an early violent merger event similar to those occurring with NGC 7252 or the Antennas, but in a lower metallicity environment. The very small number of UCDs can be explained by the fact that only a small fraction of the cluster complexes observed today appears to be gravitationally bound (e.g. Fall et al.~\cite{Fall05}, Bastian et al.~\cite{Bastia05b}). That is, only very few of the complexes will survive over Gyr time scales. Fall et al.~\cite{Fall05} argue that while between 20\% and 100\% of all stars form in clusters, the fraction of surviving clusters over a Hubble time is between 10$^{-3}$ and 10$^{-4}$ of the total stellar mass. For comparison, the total absolute brightness of all compact objects in Fornax with $M_V<-11$ mag is $M_V\simeq -15$ mag, about 6 mag or a factor of 250 fainter than NGC 1399's absolute brightness of $M_V\simeq -21$ mag (Hilker et al.~\cite{Hilker03}). This is consistent with the fractional stellar mass range for surviving clusters as estimated by Fall et al. and still leaves sufficient space for the main body of the fainter globular clusters.

It is important to stress that in order to create a consistent link between YMCs and UCDs in terms of metallicity, one would have to know the properties of an W3-like object created about 10 Gyrs ago and then interpolate up to the present. Star formation in the early universe was certainly not less intense than it is now, and mergers also happened more often than now. If the molecular clouds out of which early YMCs formed consisted of pre-processed material, one could have formed moderately metal-rich YMCs with Fornax UCD metallicity, i.e. about 1/2 to 1/4 of the metallicities derived for present YMCs, relatively early in the universe. Note that the CM trend amongst the Fornax UCDs / bright GCs (see Sect.~\ref{cmrel}) may be explained with the YMC scenario: statistically, the more massive YMCs originate from more massive progenitor galaxies and more intense star-bursts (see e.g. Elmegreen et al.~\cite{Elmegr93}) and might therefore also be more metal-enriched.\\

To finish the discussion, we consider in how far global differences between the Fornax and Virgo clusters are expected to influence the occurence of the two competing UCD formation channels. This may help to explain why properties of Virgo UCDs fit better the stripping scenario, while Fornax UCDs are more consistent with aged YMCs.

The Fornax cluster has a two times higher central galaxy density and a two times lower velocity dispersion than the Virgo cluster (Drinkwater et al.~\cite{Drinkw01}, Mieske et al.~\cite{Mieske04a}, Richtler et al.~\cite{Richtl04}, Ferguson~\cite{Fergus89}). Since the galaxy merger rate in cluster environments has the approximate dependence $n^2 \sigma^{-3}$ (Makino \& Hut~\cite{Makino97}), one expects a much higher rate of galaxy mergers in Fornax than in Virgo. This qualitative difference should also hold at the time of the most violent galaxy mergers and hence YMC formation, even if both clusters were still in their early phase of evolution and had not yet settled into virial equilibrium. If the tidal stripping scenario and the fading of YMCs created in violent galaxy mergers are competing UCD formation channels, the probability of the YMC channel is therefore higher for Fornax than for Virgo.
This fits to the observed properties of Fornax and Virgo UCDs. 

To what extent is the stripping scenario influenced by the different cluster properties? Because of the higher mass of Virgo and its correspondingly deeper potential well as compared to Fornax, Bekki et al.~\cite{Bekki03} predict the number of UCDs in Virgo created by stripping dE,Ns to be about a factor of three higher than in Fornax. 
Given that the numbers of UCDs in Virgo and Fornax (Jones et al.~\cite{Jones05} and Drinkwater et al.~\cite{Drinkw00}) do not differ by such a large factor, a plausible interpretation that fits to the observed UCD properties is that the population of Fornax UCDs is fueled by an additional formation channel.
\label{discussion}
\section{Summary and conclusions}
\label{conclusions}
\noindent We have presented spectroscopic [Fe/H] estimates for a sample of 26 compact objects in the central Fornax cluster, spanning the luminosity range $18<V<21$ ($-13.4<M_V<-10.4$) mag. Our sample includes UCDs at the bright end and is dominated by GCs at the faint end. We furthermore present [Fe/H] measurements for the nuclear regions of five Fornax dE,Ns.

We draw two main results from our measurements:

A. The mean metallicity of bright compact objects ($M_V<-11$ mag) is [Fe/H]$=-$0.62 $\pm$ 0.05 dex, which is 0.56 $\pm$ 0.15 dex higher than the mean metallicity  of $-$1.18 $\pm$ 0.15 dex for faint compact objects ($M_V>-11$ mag). The difference is significant at the 3.7$\sigma$ level. This metallicity break at $M_V=-11$ mag ($\simeq 3\times 10^6 M_{\sun}$) is accompanied by a change in the size-luminosity relation for compact objects, as deduced from ACS-imaging: for $M_V<-11$ mag, $r_h$ scales with luminosity, while for $M_V>-11$ mag, $r_h$ is luminosity-independent and of typical value for GCs. Due to these characteristic features in the metallicity and size distribution, we identify compact objects with $M_V<-11$ mag as UCDs and those with $M_V>-11$ mag as GCs. The metallicity difference between UCDs and GCs is consistent with their color difference in $(V-I)$ under the assumption of coeval populations.

B. The mean metallicity of the five investigated dE,N nuclear regions is [Fe/H]$=-$1.38 $\pm$ 0.15 dex, which is 0.75 $\pm$ 0.17 dex lower than that of the UCDs, a difference significant at 4.5$\sigma$. From the balance of two counteracting observational biases, we expect the mean metallicity of the actual nuclei to be about 0.1-0.2 dex lower, increasing the disagreement with the UCDs to at least 0.8 dex. The metallicity difference between UCDs and nuclei is slightly higher but still marginally consistent with that expected from their color difference. This implies that UCDs are coeval with or slightly younger than the nuclei. We find our metallicity estimates for UCDs close to but slightly below those derived for young massive clusters (YMCs).

From these findings and incorporating additional knowledge from the literature, we draw the following conclusions regarding the origin of UCDs:

1. We disfavor the hypothesis that most Fornax UCDs are the remnant nuclei of tidally stripped dE,Ns, given the significantly higher metallicity of UCDs compared to nuclei.

2. Although the Fornax UCDs match the metallicity and color of the red, metal-rich peak of the GC distribution, they are unlikely to be single, super-massive GCs due to their larger average sizes. Their sizes are more consistent with those predicted for merged super-clusters.

3. We propose that most Fornax UCDs are successors of merged YMCs produced in earlier star formation bursts in the course of gas rich galaxy-galaxy mergers. In contrast, for the Virgo cluster we suggest that the stripping scenario is the more important UCD formation channel. Four facts favor this distinction of formation channels:
A) The Virgo UCDs are - in contrast to the Fornax UCDs -  on average metal-poor and of comparable color to dE,N nuclei. B) Some Virgo UCDs possess high M/L ratios that may require dark matter, while Fornax UCDs have lower M/L ratios. C) The Fornax cluster has a higher current galaxy merger rate than Virgo, favoring the YMC scenario. D) The Virgo cluster has a higher mass and deeper potential well, favoring the tidal stripping scenario.\\

\begin{figure}
\plotone{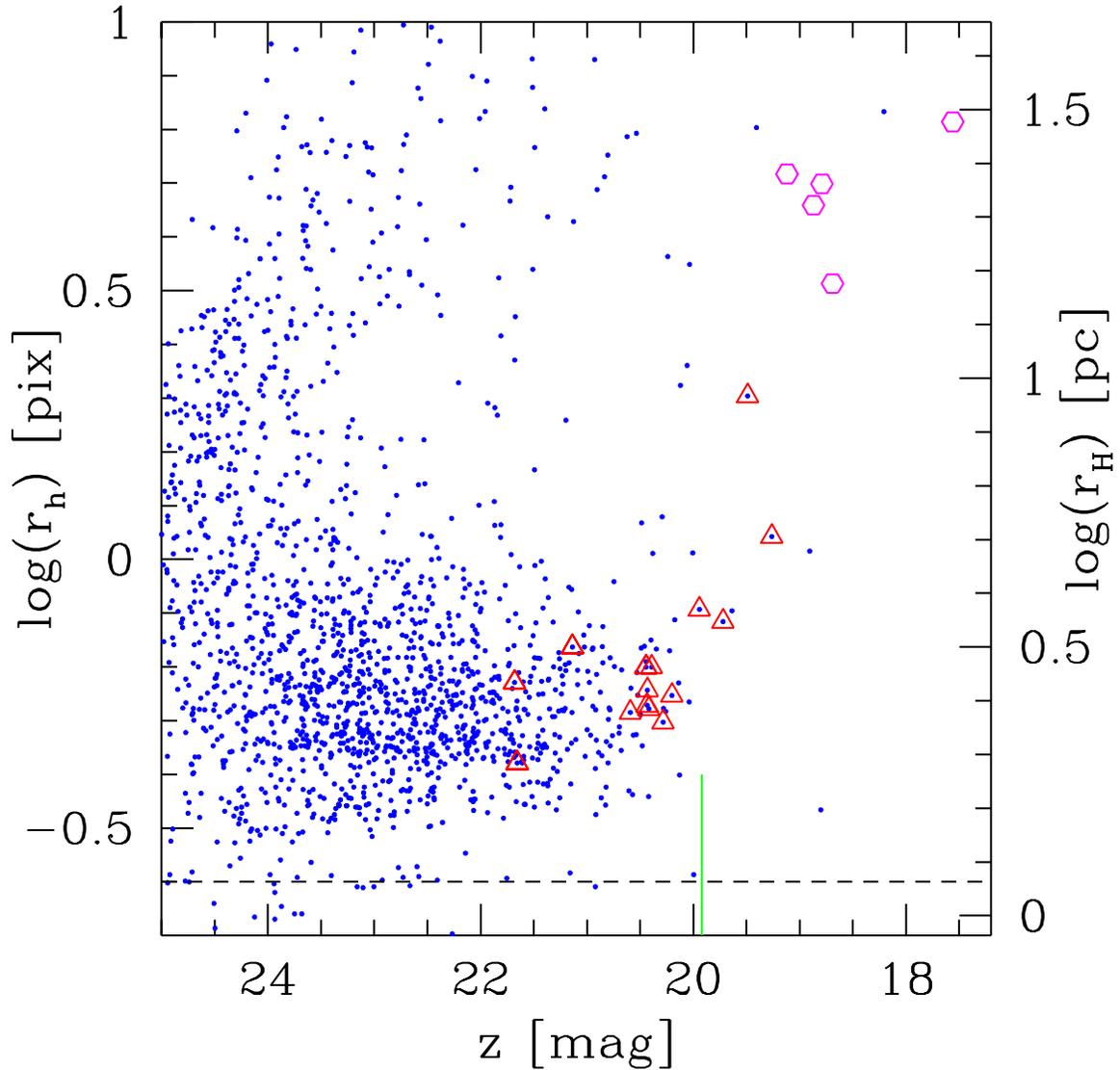}
\caption{\label{size}Triangles indicate instrumental magnitude vs. half-light radius $r_h$ for those objects from Fig.~\ref{map} with $V<21.5$ mag that were imaged in the ACS Fornax Cluster Survey FCS (Jord\'an et al.~\cite{Jordan05b} and Jord\'an et al. in preparation). These sources were contained in two of the 43 pointings of the FCS. Dots show {\it all} objects detected in these two pointings. Objects with $21<z<24$ mag and $2<r_h<3$ pc are dominated by the globular cluster system of NGC 1399. The horizontal dashed line indicates the resolution limit of the ACS data (Jord\'an et al.~\cite{Jordan05a}). Hexagons indicate the five UCDs from Drinkwater et al.~\cite{Drinkw03}, none of which was imaged in the FCS. The vertical tick at $z\simeq$ 19.9 mag marks the magnitude at which we detect the break in metallicity (see Fig.~\ref{Fe_V_nucl}). Note the coincidence of the metallicity break with the up-turn in the size distribution.}
\end{figure}

In the future, increasing the available sample of Fornax compact objects by a factor of two to three may allow us to check whether there is a {\it continuous trend} in metallicity with luminosity, as suggested from the color distribution. This will give important clues on the detailled formation histories of the Fornax UCDs.
Furthermore, one must increase the baseline for environmental comparisons of UCD properties by surveying galaxy clusters with a broad variety of intrinsic properties such as merger history, total mass and density.
\acknowledgments
\noindent We owe thanks to the staff at Las Campanas Observatory for their assistance in carrying out the observations and Gus Oemler for making the COSMOS package publicly available. We would like to thank the ACSFCS team for granting use of the data shown in Fig.~\ref{size} in advance of publication. We also thank Pat C\^ot\'e for useful discussions. SM acknowledges support by DFG project HI 855/1. LI was supported by FONDAP ``Center for Astrophysics''.
\noindent


\clearpage

\begin{deluxetable}{lccccccc}
\tablecaption{Spectroscopic and photometric properties of Fornax compact objects\label{FeHtable}}
\tabletypesize{\small}
\tablewidth{0pt}
\tablehead{
\colhead{Object-ID} & 
\colhead{$\alpha_{J2000}$} & 
\colhead{$\delta_{J2000}$} & 
\colhead{$V$} & 
\colhead{$(V-I)_0$} & 
\colhead{[Fe/H]} & 
\colhead{$H\beta$}& 
\colhead{$Mg2$}\\
\colhead{} & 
\colhead{} & 
\colhead{} & 
\colhead{(mag)} &
\colhead{(mag)} &
\colhead{[dex]} &
\colhead{[{\AA}]} &
\colhead{(mag)}}
\startdata
FCOS  1-2053 &   3:38:54.05 & -35:33:33.8 &  18.06 &   1.18 &     -0.52 $\pm$      0.11 &     2.46 $\pm$    0.09 &  0.136 $\pm$  0.022 \\
FCOS  2-2143 &   3:38:05.04 & -35:24:09.7 &  18.93 &  1.14 &     -0.69 $\pm$      0.21 &     2.32 $\pm$    0.05 &  0.084 $\pm$  0.014 \\
FCOS  1-2083 &   3:39:35.90 & -35:28:24.2 &  19.12 &  1.12 &     -0.85 $\pm$      0.29 &     2.86 $\pm$    0.12 &  0.051 $\pm$  0.023 \\
FCOS  2-2111 &   3:38:06.29 & -35:28:58.8 &  19.23 &  1.12 &      -0.90 $\pm$      0.33 &   \nodata &  0.511 $\pm$  0.016 \\
FCOS  1-021 &   3:38:41.96 & -35:33:12.9 &   19.70 &  1.16 &     -0.58 $\pm$      0.15 &     2.05 $\pm$    0.07 &  0.108 $\pm$   0.020 \\
FCOS  3-2004 &   3:39:20.50 & -35:19:14.2 &  19.73 &  1.21 &     -0.49 $\pm$      0.04 &     2.02 $\pm$    0.13 &  0.178 $\pm$  0.036 \\
FCOS  0-2031 &   3:38:28.97 & -35:22:55.9 &  19.87 &  1.11 &     -0.38 $\pm$      0.05 &     2.43 $\pm$    0.09 &   0.210 $\pm$  0.022 \\
FCOS  2-2153 &   3:38:06.48 & -35:23:03.8 &  20.04 &  1.06 &      -0.70 $\pm$      0.21 &     2.46 $\pm$     0.10 &  0.079 $\pm$  0.023 \\
FCOS  0-2066 &   3:38:23.23 & -35:20:00.6 &  20.09 &  1.10 &     -0.61 $\pm$      0.14 &     2.61 $\pm$    0.15 &  0.108 $\pm$  0.031 \\
FCOS   1-060 &   3:39:17.66 & -35:25:30.0 &  20.19 &  1.25 &     -0.59 $\pm$      0.12 &     1.86 $\pm$     0.10 &  0.136 $\pm$  0.029 \\
FCOS  0-2024 &   3:38:16.51 & -35:26:19.3 &  20.32 &  0.72 &     -0.52 $\pm$       0.10 &     1.93 $\pm$    0.11 &  0.266 $\pm$  0.034 \\
FCOS  0-2007 &   3:38:54.67 & -35:29:44.2 &  20.54 &  0.89 &       -1.00 $\pm$      0.21 &     2.81 $\pm$    0.21 &  0.174 $\pm$  0.042 \\
FCOS  2-2165 &   3:37:28.22 & -35:21:23.0 &  20.57 &  0.95 &      -0.40 $\pm$      0.05 &  \nodata &  0.176 $\pm$  0.031 \\
FCOS  0-2023 &   3:38:12.70 & -35:28:57.0 &  20.66 &  1.09 &     -1.13 $\pm$      0.14 &     3.49 $\pm$    0.15 &  0.067 $\pm$  0.025 \\
FCOS  0-2069 &   3:38:26.71 & -35:30:07.2 &  20.67 &  0.97 &      -1.50 $\pm$       0.20 &     1.56 $\pm$    0.14 &  0.027 $\pm$  0.046 \\
FCOS  1-058 &   3:38:39.39 & -35:27:06.4 &  20.67 &  1.02 &     -0.74 $\pm$      0.25 &     2.07 $\pm$    0.13 &  0.065 $\pm$  0.034 \\
FCOS  0-2072 &   3:38:32.06 & -35:28:12.7 &  20.71 &  1.14 &      -0.50 $\pm$      0.08 &     2.88 $\pm$    0.27 &  0.148 $\pm$  0.047 \\
FCOS  2-2072 &   3:38:14.69 & -35:33:40.7 &  20.72 &  0.89 &      -0.60 $\pm$       0.10 &     2.69 $\pm$    0.19 &  0.127 $\pm$   0.040 \\
FCOS  0-2041 &   3:38:44.11 & -35:19:01.6 &  20.73 &  1.30 &     -1.32 $\pm$      0.25 &     3.33 $\pm$    0.33 &  0.052 $\pm$  0.051 \\
FCOS  0-2032 &   3:38:30.22 & -35:21:31.0 &   20.80 &  0.84 &     -1.75 $\pm$       0.30 &     3.14 $\pm$    0.68 &  0.055 $\pm$  0.129 \\
FCOS  2-086 &   3:37:46.77 & -35:34:41.7 &  20.81 &  0.90 &     -0.66 $\pm$      0.17 &     3.05 $\pm$    0.23 &  0.242 $\pm$  0.041 \\
FCOS  1-2115 &   3:38:49.18 & -35:21:42.1 &  20.85 &  1.20 &     -1.26 $\pm$      0.25 &      2.50 $\pm$    0.18 &  0.491 $\pm$  0.039 \\
FCOS  2-089 &   3:38:14.02 & -35:29:43.0 &  20.87 &  1.06 &       -2.00 $\pm$      0.36 &    \nodata &  0.015 $\pm$  0.044 \\
FCOS  0-2074 &   3:38:35.66 & -35:27:15.5 &  20.91 &  1.04 &     -2.25 $\pm$      0.27 &     2.17 $\pm$    0.12 & -0.006 $\pm$  0.033 \\
FCOS    2-2100 &   3:38:00.17 & -35:30:08.3 &  20.94 &  1.05 &     -0.91 $\pm$      0.24 &     1.91 $\pm$    0.16 &  0.069 $\pm$  0.046 \\
FCOS  0-2092 &   3:39:05.02 & -35:26:53.9 &  20.96 &  1.13 &     -1.66 $\pm$      0.38 &     1.59 $\pm$    0.14 & -0.006 $\pm$  0.047 \\
\enddata
\tablecomments{ Table showing the [Fe/H], H$\beta$ and $Mg2$ values for those 26 out of the 35 observed compact objects in Fornax with sufficiently high S/N. The first line gives the FCOS identifier (see Mieske et al.~\cite{Mieske04a}). Photometry is from Mieske et al.~\cite{Mieske04a}. Original $(V-I)$ values have been dereddened using Schlegel et al.~\cite{Schleg98}. Note that for sources FCOS 0-2xxx, $V$ and $(V-I)_0$ are
derived indirectly from the $C-R$ photometry of Dirsch et al.~\cite{Dirsch03}, see Mieske et al.~\cite{Mieske04a}. ``\nodata'' in the H$\beta$ column means that the index feature fell into a chip gap.}
\end{deluxetable}
\end{document}